\documentclass[conference]{IEEEtran}
\IEEEoverridecommandlockouts
\usepackage{cite}
\usepackage{amsmath,amssymb,amsfonts}
\usepackage{algorithmic}
\usepackage{graphicx}
\usepackage{textcomp}
\def\BibTeX{{\rm B\kern-.05em{\sc i\kern-.025em b}\kern-.08em
    T\kern-.1667em\lower.7ex\hbox{E}\kern-.125emX}}
\begin{document}

\title{Data-Driven and Deep Learning Methodology for Deceptive Advertising and Phone Scams Detection}

\author{$^\dagger$TonTon Hsien-De~Huang,
        $^*$Chia-Mu~Yu,
        and~$^*$$^\dagger$Hung-Yu Kao\\
$^\dagger$Leopard Mobile Inc.\\
$^*$Department of Computer Science and Engineering, National Chung Hsing University, Taiwan\\
$^\dagger$ $^*$$^\dagger$Department of Computer Science and Information Engineering, National Cheng Kung University, Taiwan
}

\maketitle

\begin{abstract}
The advance of smartphones and cellular networks boosts the need of mobile advertising and targeted marketing. However, it also triggers the unseen security threats. We found that the \emph{phone scams} with fake calling numbers of very short lifetime are increasingly popular and have been used to trick the users. The harm is worldwide. On the other hand, \emph{deceptive advertising} (\emph{deceptive ads}), the fake ads that tricks users to install unnecessary apps via either alluring or daunting texts and pictures, is an emerging threat that seriously harms the reputation of the advertiser. To counter against these two new threats, the conventional blacklist (or whitelist) approach and the machine learning approach with predefined features have been proven useless. Nevertheless, due to the success of deep learning in developing the highly intelligent program, our system can efficiently and effectively detect phone scams and deceptive ads by taking advantage of our unified framework on \emph{deep neural network (DNN)} and \emph{convolutional neural network (CNN)}. The proposed system has been deployed for operational use and the experimental results proved the effectiveness of our proposed system. Furthermore, we keep our research results and release experiment material on http://DeceptiveAds.TWMAN.ORG and http://PhoneScams.TWMAN.ORG if there is any update.
\end{abstract}


\section{Introduction}
\begin{figure}[hbtp]
	\centering
	\includegraphics[width=3in,height=2.8in]{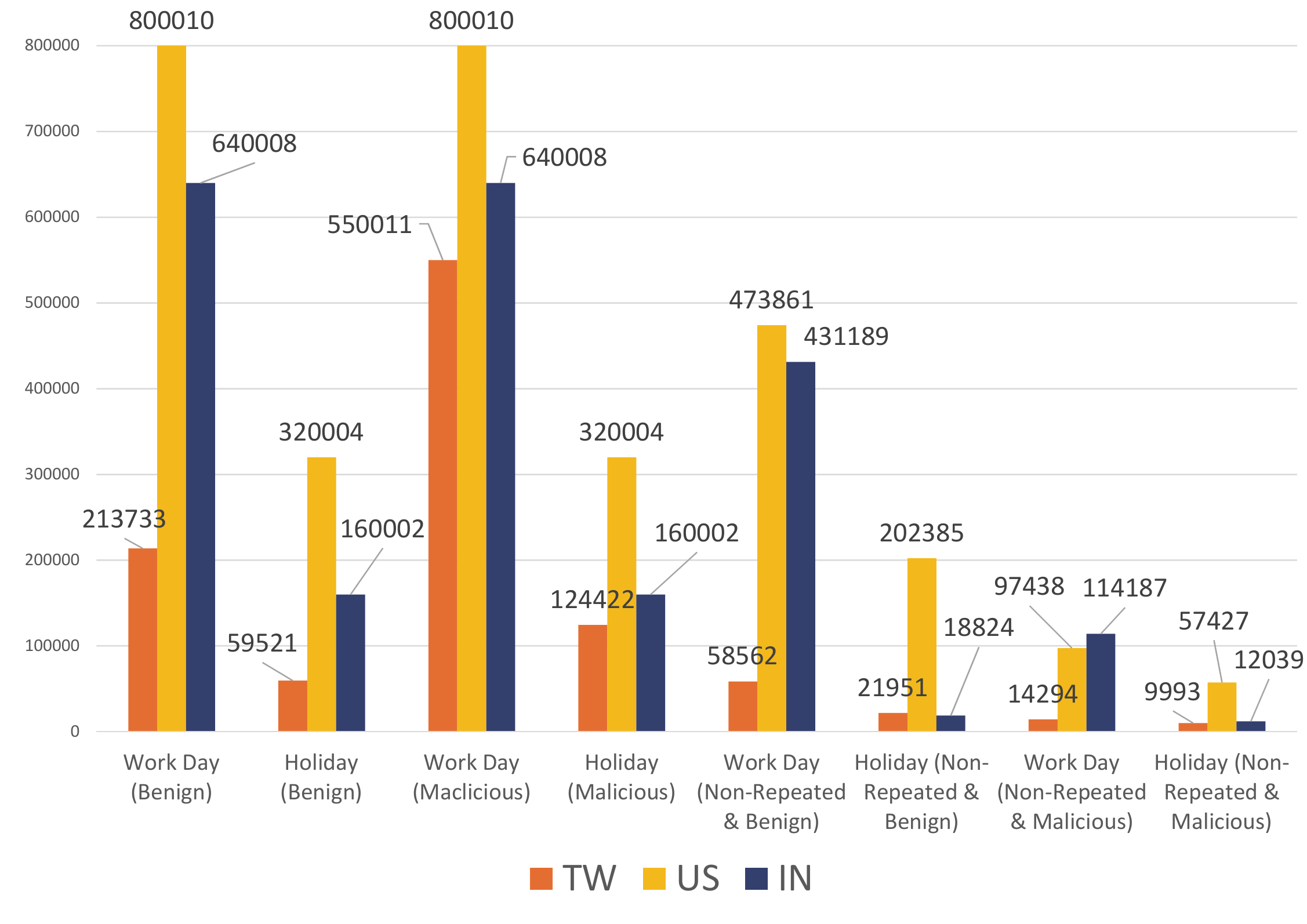}
	\caption{The current situation of the phone scams analysis.}\label{fig: F001}
\end{figure}

Federal Trade Commission (FTC) indicated that phone scam is the most popular type of scams in United States and there are more than 150 million of disputes. In particular, in 2014, the amount of monetary loss from phone scam is more than 1.7 billion. 46 percent of victims can clearly indicate how they are tricked but 54 percent of victims claimed to be tricked by phone scam. At the same time, the strike against robocalls by FTC is supported by primary companies such as AT\&T and Google etc. On the other hand, there are cases where the college students in China are tricked to give away their tuition fees. According to the report, there are approximately 1.6 million of people conducting scam business and the revenue of this business is more than 110 billion RMB. In addition, there are similar cases in Japan where the amount of monetary loss is more than 50 billion Yen. As shown in Fig. \ref{fig: F001}, in the case of non-repeated calling numbers, all of the phone scams intensively occur during the weekdays. More specifically, compared to such an intensive amount of phone scams during the weekdays, only approximately one third of phone scams occurs during the weekend in the United States, India, and Taiwan. In the case of repeated calling numbers, the situation remains unchanged; very little portion of phone scams occurs during the weekend. 

The Ministry of Public Security, China, reported in 2016 that phone scams should account for the significant financial losses (more than 2M RMB). In addition, recently each online user receives 21.3 harassing calls in average per week. Among them, $20\%$ of harassing calls are dialed by Internet phone or fake numbers. However, this occupies $60\%$ of the total financial loss, reported by the Internet Society of China. Such a large number of phone scams can be attributed to the fact of the massive leakage of personal information. Thus, FTC announced a list of dialing numbers for reference; once the call is from those numbers, it is likely to be a harassing call. Nonetheless, the reality is far more complicated; for example, the list announced by FTC is not working for the call from non-US area. Actually, phone scams can be categorized as follows \cite{ftc}:
\begin{enumerate}
  \item Free Vacations and Prizes
  \item Loan Scams
  \item Phony Debt Collectors
  \item Fake Charities
  \item Medical Alert/Scams
  \item Targeting Seniors
  \item Warrant Threats
  \item IRS Calls
\end{enumerate}

\begin{figure}[hbtp]
	\centering
	\includegraphics[width=3in,height=2.1in]{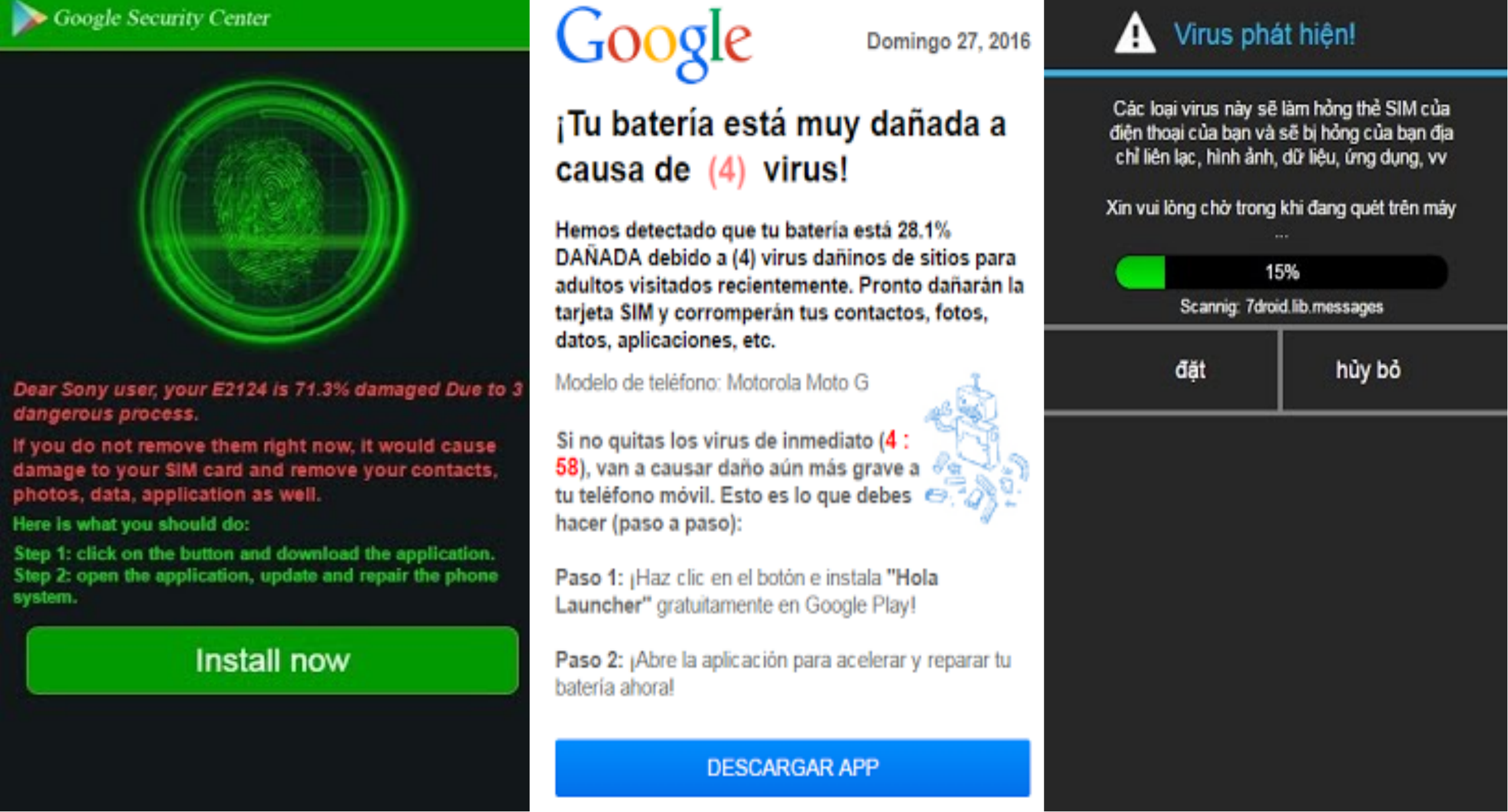}
	\caption{Examples of deceptive ads.}\label{fig: F002}
\end{figure}

When you are doing web surfing, you might have experience on being informed that your cellphone is infected by malware/virus or on being forced to download unknown Apps. Deceptive ads, a particular type of ads, is the use of false or misleading statements in advertising, and may cause negative impact on stakeholders. Deceptive ads is a particular type of malicious advertising \cite{lzxyw12} that tricks you to download unnecessary Apps. Deceptive ads might give you false alarms, saying “you are infected” or “your cellphone needs to be updated”. These false alarms sometimes cannot be turned off by refreshing the webpage, making you uncomfortable with the status of the cellphone. Such deceptive ads is pretty annoying; the advertisers may in turn gain reverse effect (e.g., awful reputation) by adopting deceptive ads. The detection and even capture of deceptive ads is definitely not easy; the contents and wordings varied in different regions, time zones, and countries. Phone scams and deceptive ads are not a new threat; however, since mobile devices are always on and carried 24/7, such a threat has quickly ascended on the list of everyday Internet threats. Deceptive ads exhibit fast-flux behavior, while phone scams exhibit caller ID spoofing behavior and therefore is more difficult to be caught. Fig. \ref{fig: F002} shows examples of deceptive ads whose pop-up ads show falsified information and instructions.

\begin{figure}[hbtp]
	\centering
	\includegraphics[width=3in,height=2.1in]{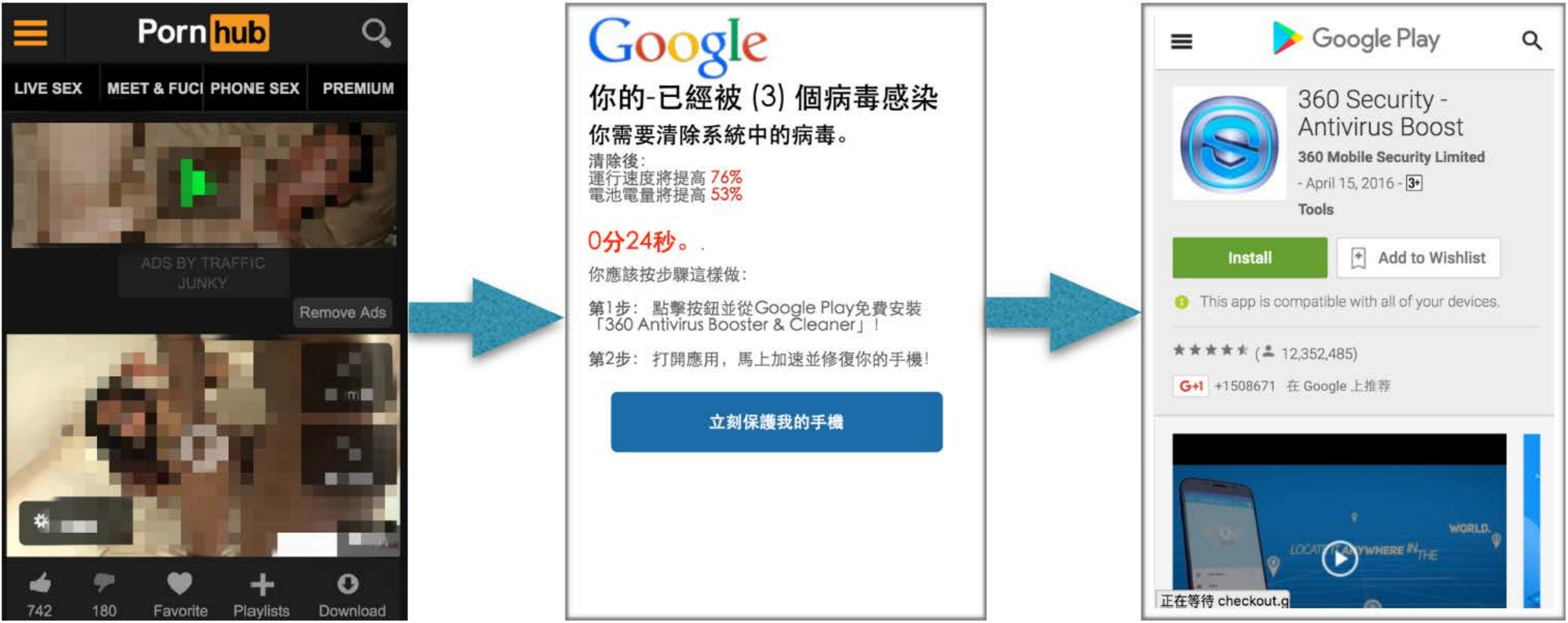}
	\caption{Procedures of deceptive ads.}\label{fig: F003}
\end{figure}

\begin{figure}[hbtp]
	\centering
	\includegraphics[width=3in,height=1.3in]{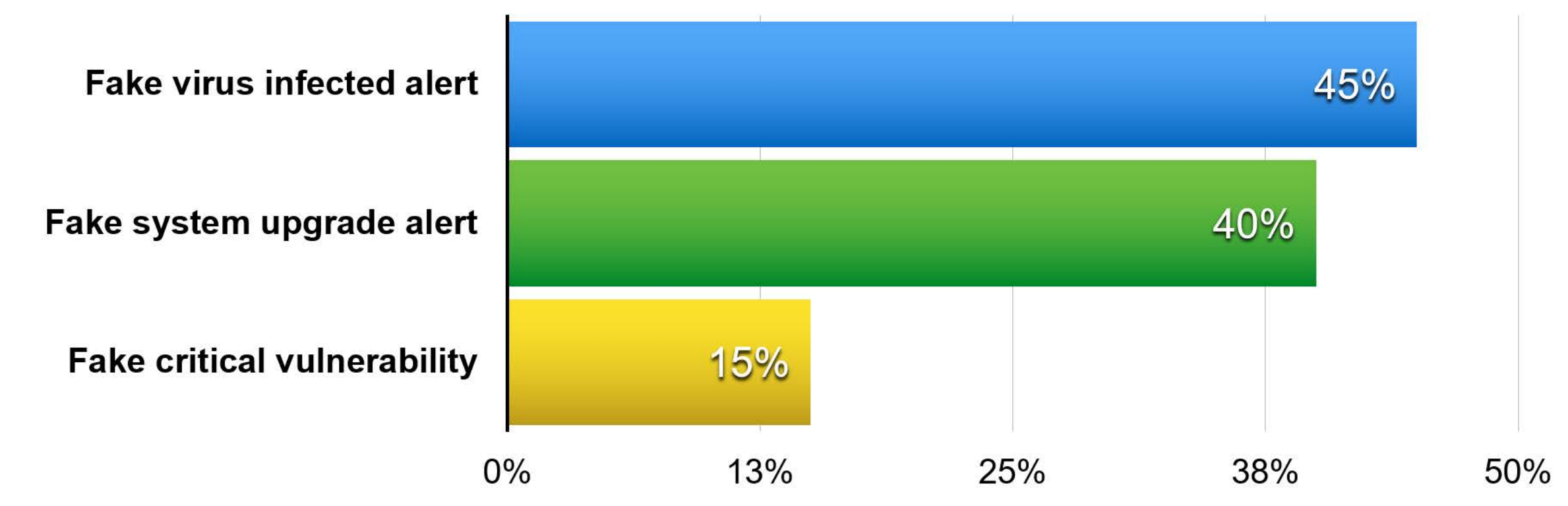}
	\caption{Three categories of deceptive ads.}\label{fig: F004}
\end{figure}

On the other hand, we found that deceptive ads is increasing its impact on cellphone. In general, the typical procedures of how deceptive ads work are shown as follows (see Fig. \ref{fig: F003}). 
\begin{enumerate}
  \item Browse the porn website and click the temptation ads.
  \item Pop up a false or misleading statement.
  \item Orientate user towards the anti-virus App on Google Play.
\end{enumerate}	
Among the deceptive ads, three main categories in terms of the contents will be displayed (see Fig. \ref{fig: F004}).
\begin{itemize}
  \item Pretend to have an alert for malware/virus infection.
  \item Pretend to have a notification for updating Android system.
  \item Pretend to have a notification for patching the critical vulnerability.
\end{itemize} 
Note that among the three categories, fake infection alert occupies almost the half. Only a few deceptive ads are the fake patching notification. Note also that through the analysis of advertising alliance traffic of April 2016, we discovered that deceptive ads occupy up to 21\% in the global advertising alliance traffic.

Google is one of the biggest victims from the false and misleading content’s perspective. In essence, the tool Apps, such as security and power-saving Apps, suffers the most, in terms of App categories. Most of the deceptive ads falsely warn the user about malware/virus infection or battery dying. An observation from our collected ads is that the targets of deceptive ads are mostly emerging companies. An explanation to such an observation is that the deceptive ads are usually associated with the tool Apps. The emerging companies such as Qihoo 360 and Cheetah Mobile have developed and published a number of tool Apps as strategic software for expanding the market, and then deceptive ads have more impact on emerging companies. As a result, the investment of the advertising will turn to have negative impact on the reputation. Based on our study, we find that the porn websites are the major sources of the deceptive ads. Of course, the rest of the deceptive ads are mainly from the game portal, file hosting service, and news websites. The typical scenario is that a banner showing an alert about the malware/virus infection jumps out, once the user clicks the ads, it will be directed to Google Play. The user may click to check and then will be directed to download the unnecessary App (see Fig. 3).

\section{Deep Learning Approach for Problem Solving}

Deep learning allows computational models that are composed of multiple processing layers to learn representations of data with multiple levels of abstraction \cite{dplg2015}. Recently, Alpha Go from Google DeepMind gains a huge success in Computer Go. The deep learning behind the Alpha Go receives huge attentions from both publics and academics \cite{deepmind16}. Neural network mimics how human brain works and can apply to perform pattern recognition. Therefore, the image can be input to the neural network and the burdensome procedures such as feature extraction and data model can be omitted. Based on the convolutional layer and pooling layey, CNN strenghtens the effectiveness of pattern recognition. The most well-known convolutional neural network (CNN) models are AlexNet \cite{AlexNet}, VGG \cite{VGG}, GoogleNet \cite{GoogleNet}, and ResNet in ImageNet Large Scale Visual Recognition Challenge (ILSVRC). Due to the use of deceptive ads, the existence of bad actors may lead to the vicious competition, and therefore the benign actors haves less earning. We observe that the deceptive ads mutate very frequently, almost in real time. Thus, in the case of no dedicated real-time detection, it is very difficult to promise the contents of the ads. 

The technical challenge in differentiating between benign and deceptive ads is that the contents of deceptive ads is not fixed; instead, the contents of deceptive ads vary according to the factors such as the region, time zone, language, etc. The conventional detections are useless in detecting deceptive ads. Our previous research results have adopted machine learning approach (with automated feature extraction for further text-mining and pattern recognitions to handle multilingual deceptive ads) and integrated it with techniques including Ontology, Type-2 Fuzzy Logic etc. However, our previous mechanism was unable to smoothly handle these fast flux-like behaviors as above technical challenge \cite {huangsp2016}. Thus, our goal is to increase expressiveness of our mechanism, while smoothly handling these fast flux-like behaviors of the deceptive ads challenge. 

The most frequently used method of detecting deceptive behavior includes phone scams and deceptive ads. There have only been few research efforts for yellow pages, blacklist, detecting malicious ads and phishing on malicious URLs. For example, the pre-defined features or fixed delimiters for feature selection and reactive URL or phone number blacklisting can be used for the detection \cite{phishstorm} \cite{maliciousurl}. However, these techniques are inefficient in detecting phone scams or deceptive ads due to the short lifetime of deceptive behavior. In fact, the most common type of phone scams is “fake phone number” which makes phone calls from outside and deceptive ads exhibit a fast flux-like and region-aware behaviors. The former means that the phone number exhibit caller ID spoofing behavior, while the latter means that the deceptive ads would be valid for only a very short time interval and by visiting the same deceptive ads URL, the browsers in different countries or even the same browser with different language settings may see different contents\footnote{https://www.youtube.com/watch?v=qxqvtKrrVBA}\footnote{https://www.youtube.com/watch?v=16nvbkUdX6k} (see Fig. \ref{fig: F002}). These two features make tracking down the deceptive behavior much more difficult.

\section{Our Proposed System}

Our system try to combine the novel techniques on user’s cellphone (client-side), and deep learning, a branch of machine learning based on a set of algorithms that attempt to model high-level abstractions in data by using multiple processing layers, with complex structures or otherwise, composed of multiple non-linear transformations, on our back-end system (server-side) to address the challenge of accurately detecting phone scams for the loss of money and deceptive ads for the product reputation, and the ads cost saving. Based on the DNN and CNN in deep learning, we proposed a mechanism to detect phone scams and deceptive ads. Our system is a cloud-assisted host-based detection. One may see that once the user device (e.g., smartphone) is reached by a unknown people or goes to an URL, the encoded or hashed information about the browsing (e.g., URL) will be sent to the cloud service implemented with our detection core. Afterwards, the cloud will return the detection result to the user device. Our mechanism is featured by both the client-side real-time lightweight defense and the cloud-side deep inspection. Such a hybrid detection mechanism aims to detect the deceptive ads with significantly low false positives and false negatives. As the first step toward the detection, our system is implemented by Python language and integrated with some open source projects such as asynchronous and customizable analysis platform: IRMA\footnote{http://irma.quarkslab.com}, Elasticsearch which provides the most powerful full-text search capabilities\footnote{https://www.elastic.co} and Kibana which easily visualizes data\footnote{https://www.elastic.co/products/kibana}. We also provide a simple RESTful API using JSON formatted report over HTTP.

Here, the \emph{unknown calling data} refers to the calling data contributed only by the strangers. In our system implementation, for detecting phone scams, first we collect a huge set of unknown calling data from user’s feedback, based on our user-base and back-end system. We collected 1500 thousands of unknown calling data in July, 2016 from countries such as Brasil, Italy, UK and France etc., and then captured the feature as the input to our system to learn the difference between benign and deceptive phone numbers. 

\begin{figure}[hbtp]
	\centering
	\includegraphics[width=3in,height=2.55in]{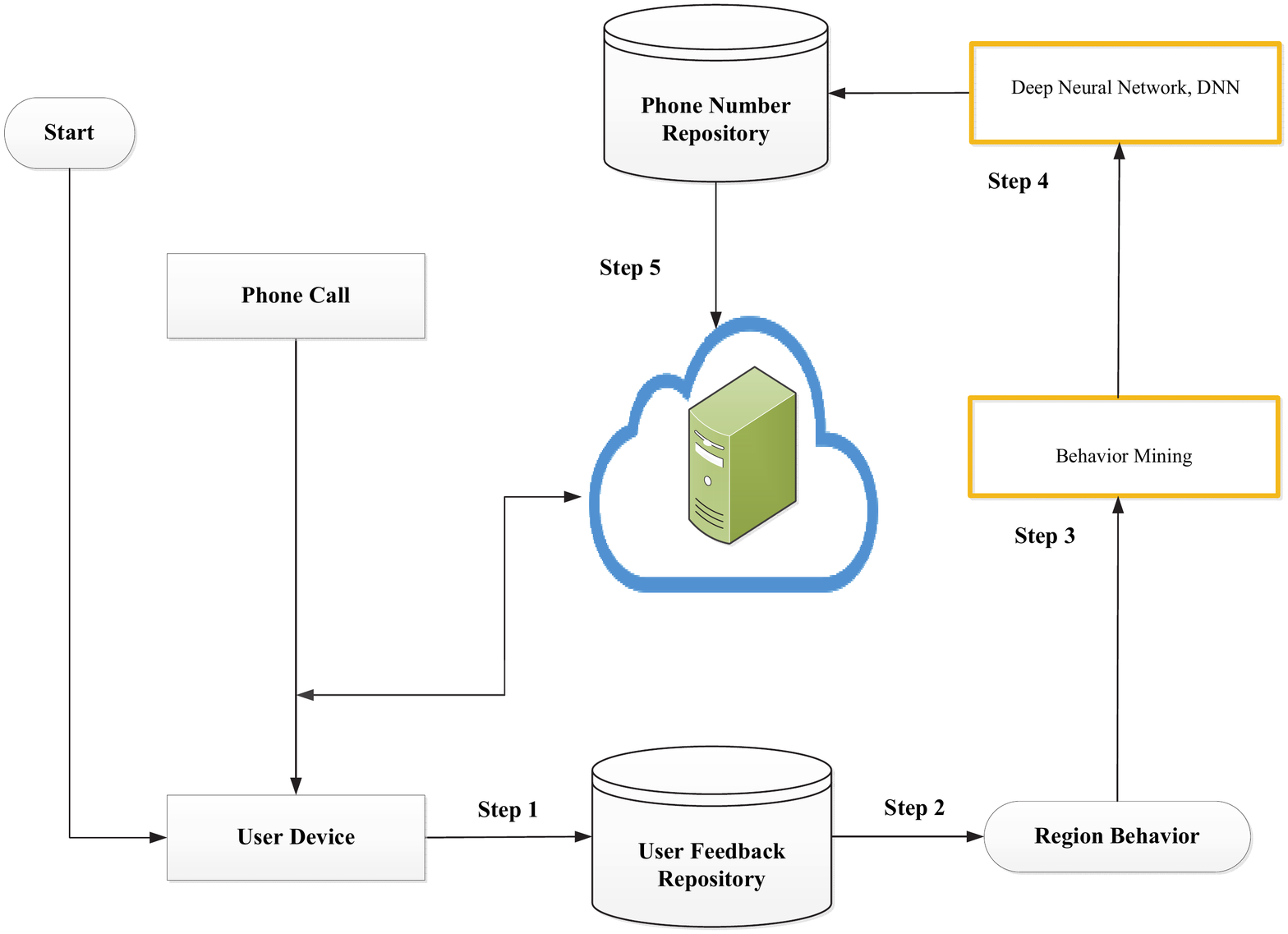}
	\caption{The architecture of our phone scams detection system.}\label{fig: F005}
\end{figure}

More specifically, as shown in Fig. \ref{fig: F005}, the user received the call from the strangers. Our App on the user’s cellphone sent a feedback to the back-end system (step 1). On the way to the back-end system, an online database with regional data was checked. In our mechanism, the aim of doing so is to trigger the region-specific and country-specific phone scams, with the observation that certain phone scams may check where the user is from and shows different behavior (step 2). And then, our mechanism extracted the time length of ringing, speaking time, user action, hanging out, picking up, timeout, blacklist, and calling time etc. from the phone numbers records as features and then feed to our DNN system (step 3). Afterwards, we obtain the classification results from DNN (step 4, via Google’s TensorFlow \cite{tensorflow}). 

More precisely, all of the phone numbers records' feature from the unknown calling data were input to neural network and the layers of neural network learned the best results. The neural network ran many times until the overall error rate reached the minimum. Thus, in the records of phone number, deep learning helped us find the results for the classification. After that, both the cloud-side system and client-side defense were updated (step 5). A brief description of our model is shown as followings:

\begin{enumerate}
  \item the activation function is relu.
  \item for the first layer, the number of input neurons is $9$ while the number of output neurons is $14$.
  \item for the second layer, the number of input neurons is $14$ while the number of output neurons is $9$.
  \item for the third layer, the number of input neurons is $9$ while the number of output neurons is $5$.
  \item the last layer is the output layer with the sigmoid activation function.
\end{enumerate}

\begin{figure}
	\centering
	\includegraphics[width=3in,height=2.55in]{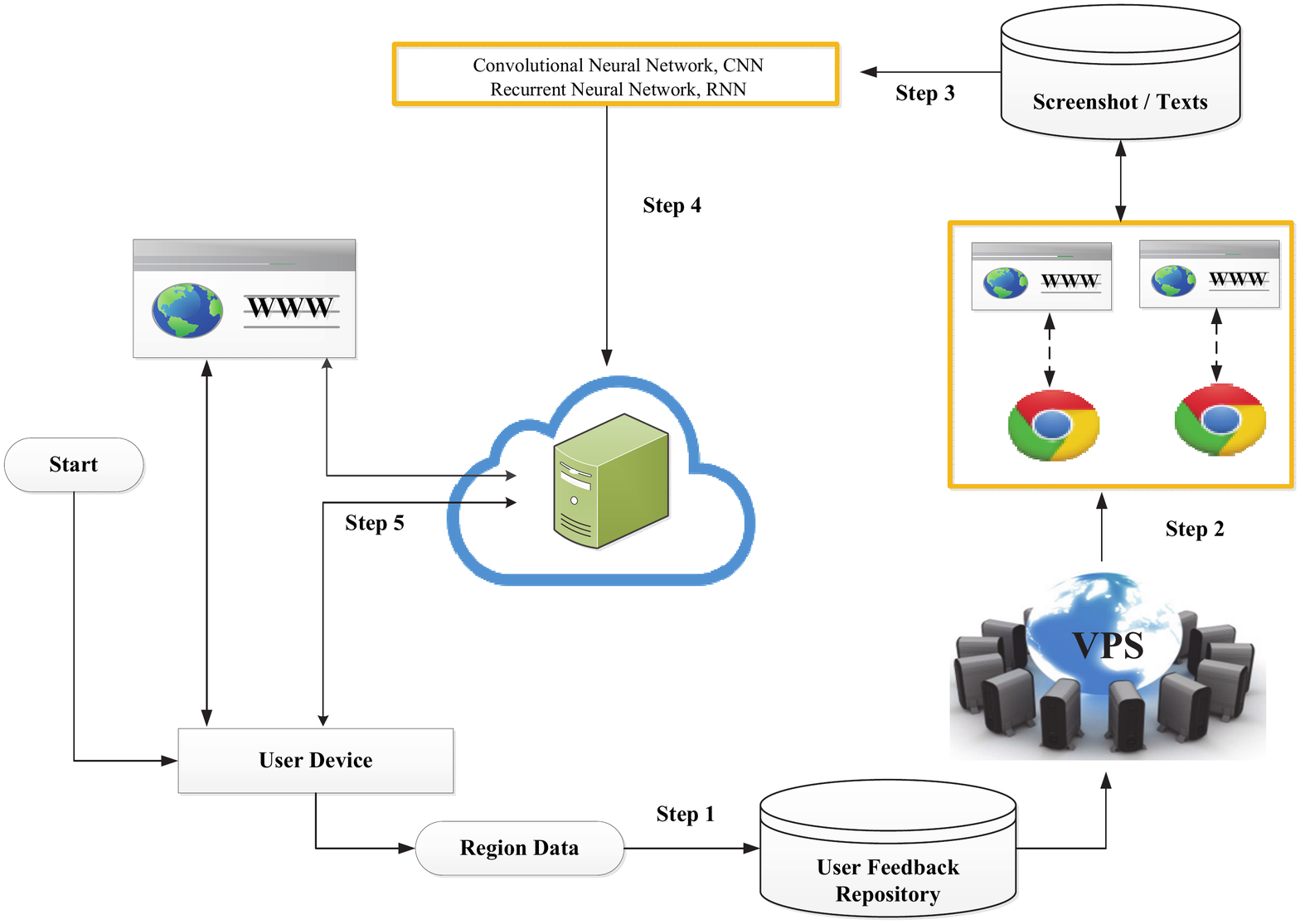}
	\caption{The architecture of our Ads detection system.}\label{fig: F006}
\end{figure}

\begin{figure}
	\centering
	\includegraphics[width=3in,height=2.3in]{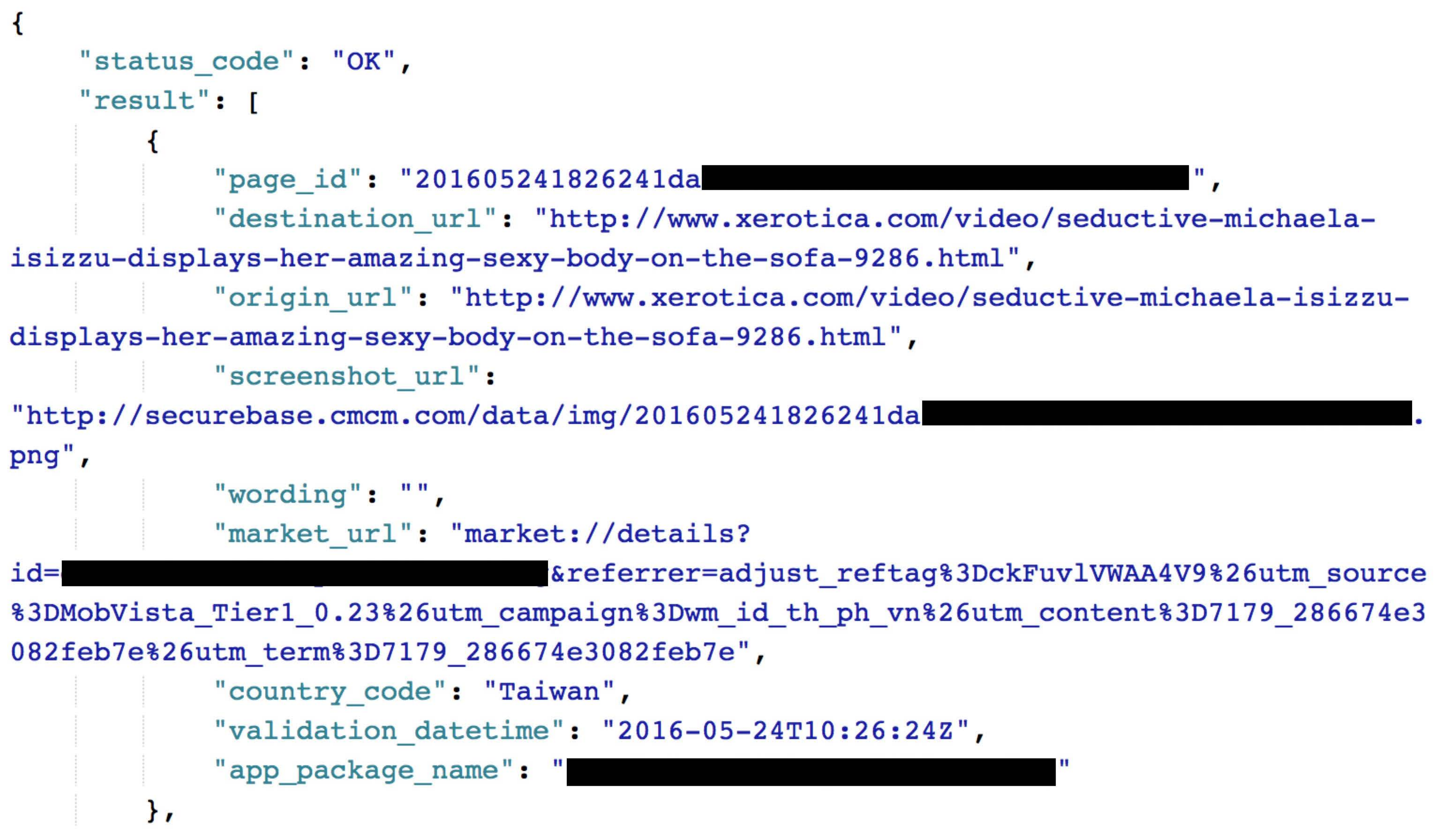}
	\caption{Our deceptive ads analysis report.}\label{fig: F007}
\end{figure}

On the other hand, for detecting deceptive ads, first we collected a huge number of ads from the Internet, based on our user base and back-end system. We collected more than 150 thousand advertising URLs daily, and then captured the advertising figures and texts as the input to our system to learn the difference between benign and deceptive ads. More specifically, as shown in Fig. \ref{fig: F006}, the user browsed the web-pages. Our App on the user’s cellphone sent a feedback to the back-end system (step 1). On the way to the back-end system, an online database with regional data were checked. In our mechanism, we deployed virtual private servers (VPSs) around the world. The aim of doing so is to trigger the region-specific and country-specific deceptive ads, with the observation that certain deceptive ads may check where the clicking user is from and shows different contents (step 2). After the VPS, our mechanism extracted the figures, texts, and HTML source file from the ads for deep learning (step 3). Afterwards, we obtained the classification results from deep learning (re-train and fine-tune Google's Inception-V3 model\footnote{. It is trained for the ImageNet Large Visual Recognition Challenge using the data from 2012. AlexNet achieved by setting a top-5 error rate of 15.3\% on the 2012 validation data set; BN-Inception-v2 achieved 6.66\%; Inception-v3 reaches 3.46\%} via TensorFlow \cite{tensorflow}). More precisely, all of images are input to neural network and the layers of neural network may automatically generate the best features, instead of manually extracting features. The neural network ran many times until the overall error rate reached the minimum (step 4). In the structure of our mechanism, one can see that more convolutional layers implied more capabilities of recognizing the object in an abstract sense. Thus, in the context of ads, deep learning helps us find the features for the classification. After that, both the cloud-side system and client-side defense were updated (step 5). In addition, we also offered RESTful API for checking whether deceptive ads cause economic impacts on their business that showed how destination URL, original URL, screenshot and date and time helped in detection, as shown in Fig. \ref{fig: F007}.

\section{Evaluation Results}

\begin{figure}[hbtp]
	\centering
	\includegraphics[width=3in,height=2.2in]{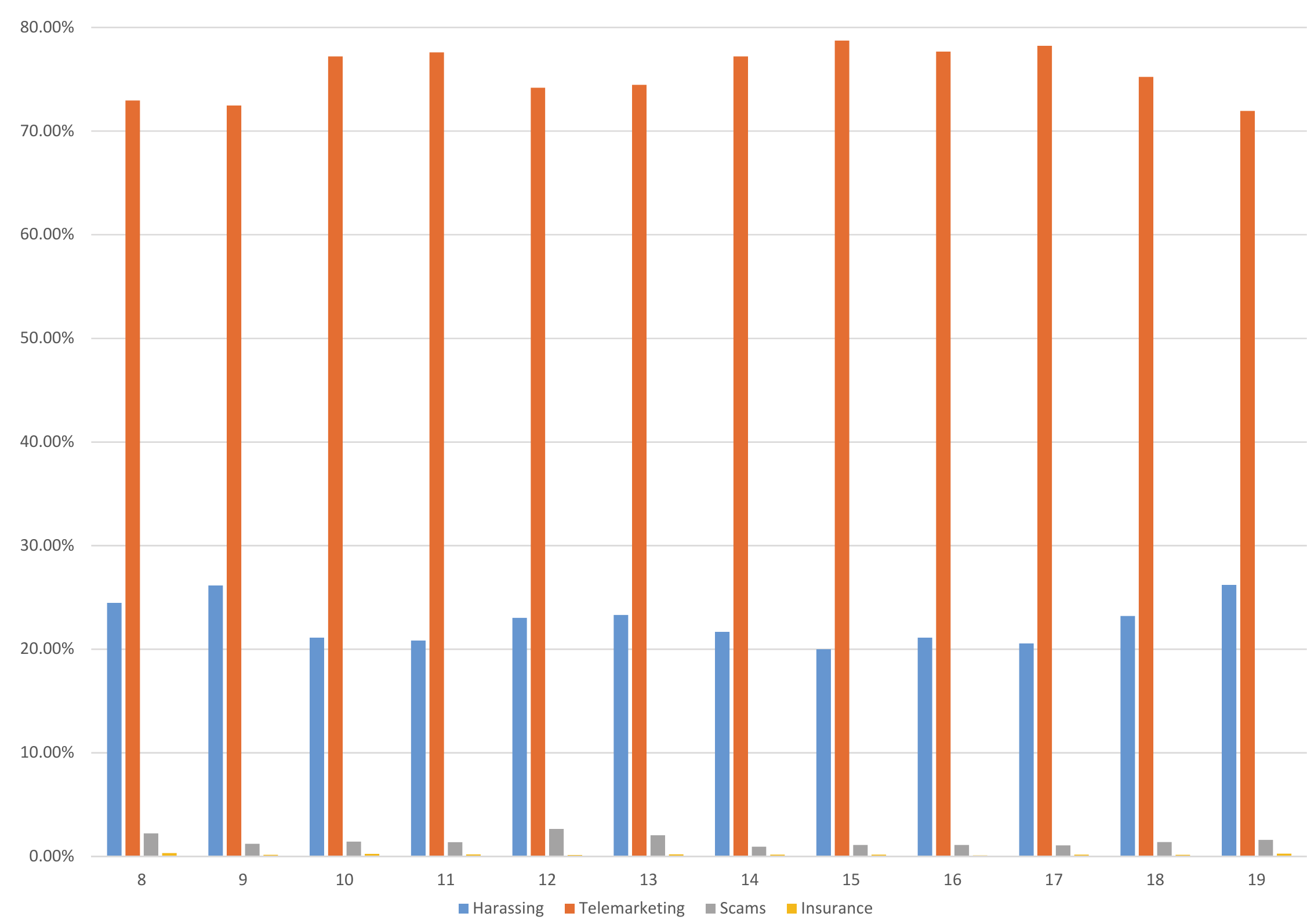}
	\caption{The average of collection of phone numbers in 1 day.}\label{fig: F008}
\end{figure}

\begin{figure}[hbtp]
	\centering
	\includegraphics[width=3in,height=2.2in]{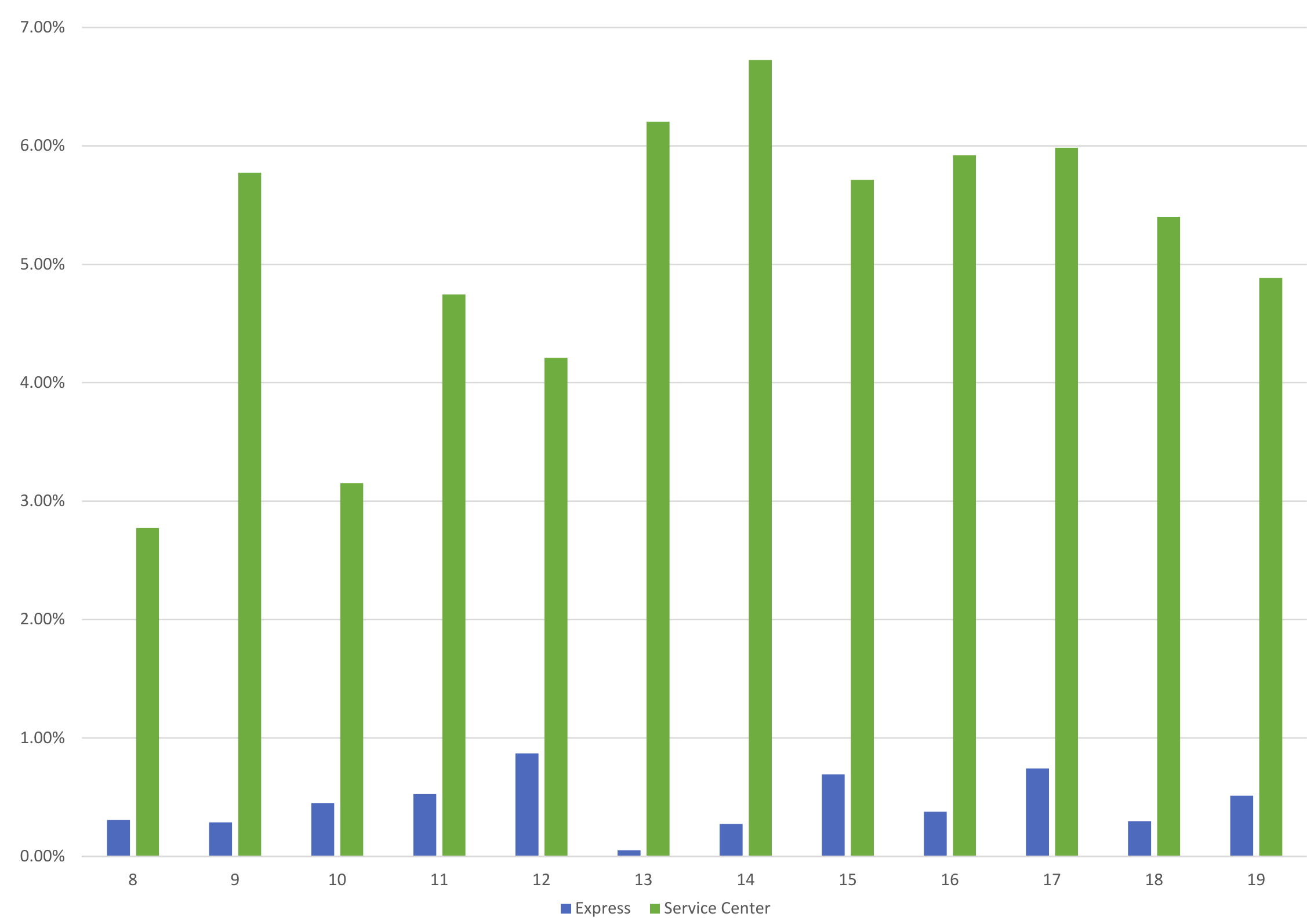}
	\caption{The average of collection of phone numbers in 1 day.}\label{fig: F009}
\end{figure}

\begin{figure}[hbtp]
	\centering
	\includegraphics[width=3in,height=2.2in]{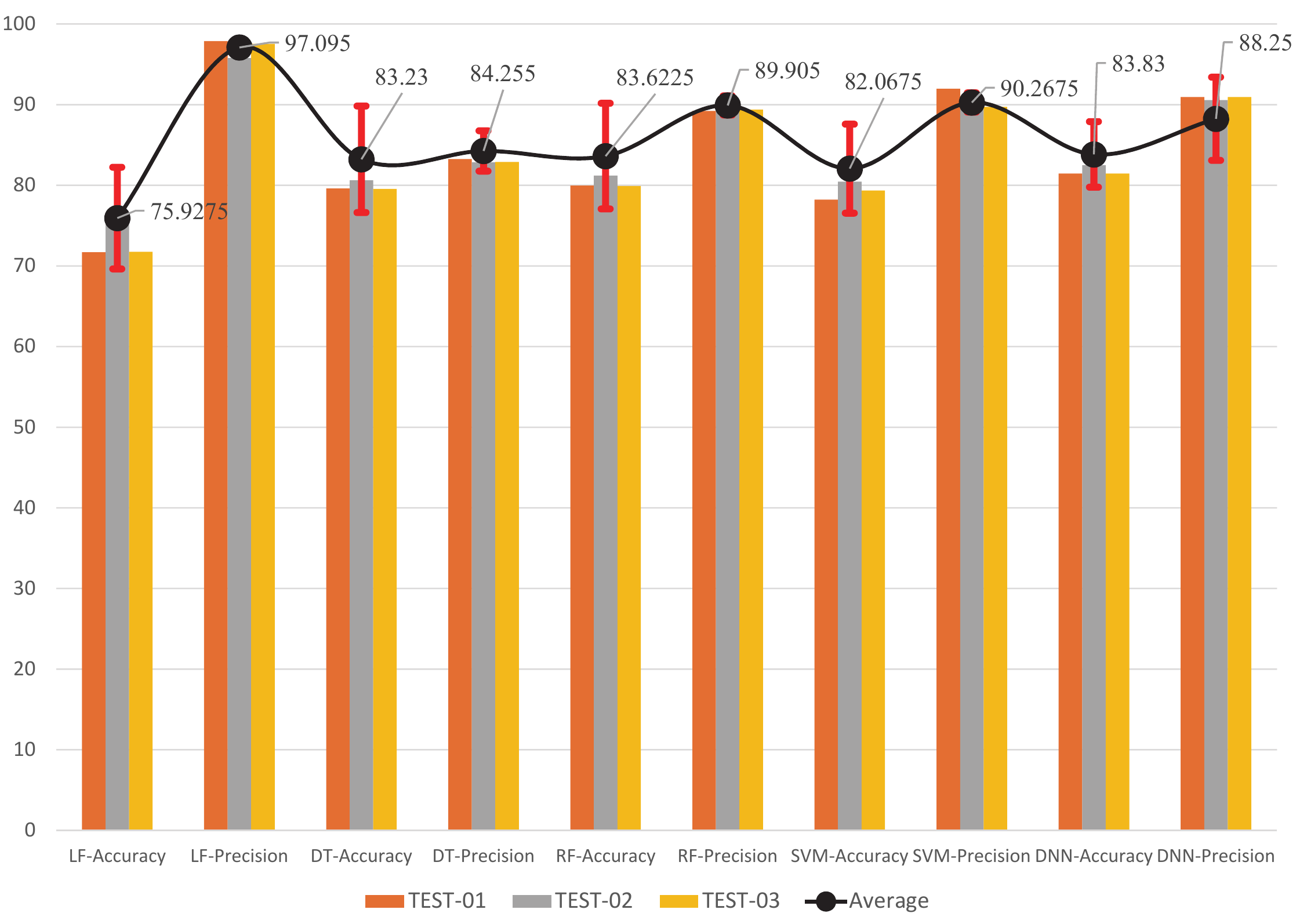}
	\caption{Our mini experiment results.}\label{fig: F010}
\end{figure}

\begin{figure}[hbtp]
	\centering
	\includegraphics[width=3in,height=2.2in]{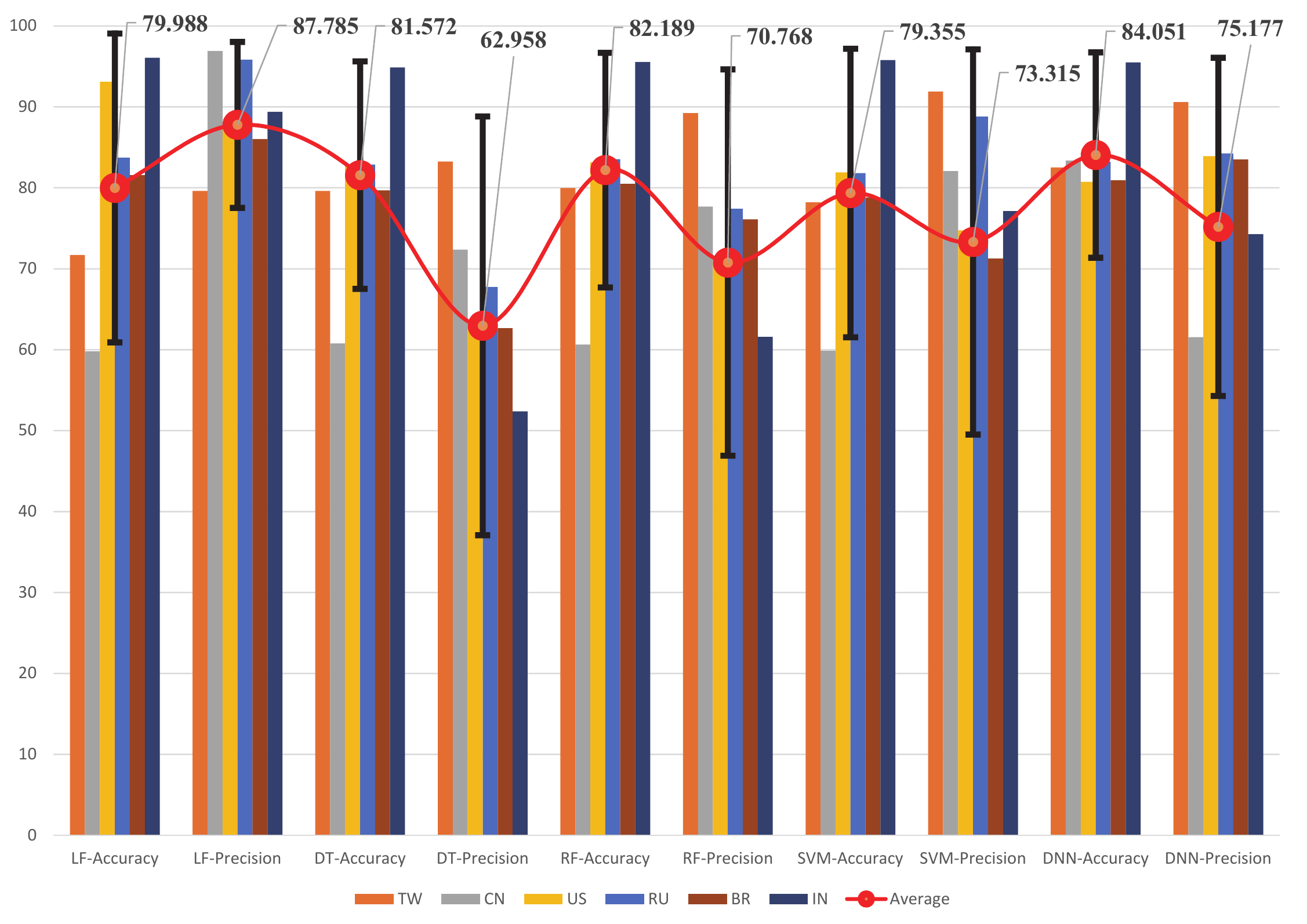}
	\caption{Our experiment results.}\label{fig: F011}
\end{figure}

Our featured products have reached 38 billion installations globally with 623 million monthly active users each month by December, 2016. For phone scams, our core product has 80 million daily active users, and 23 million daily active users have received phone call from strangers including 10\% phone scams. Fig. \ref{fig: F008} and Fig. \ref{fig: F009} show the average of collection of phone numbers in one day. We found that during the working time, the so-called malicious calling such as harassing, telemarketing, scams and insurance could be a huge portion. We also found that the normal calling from, for example, express, service center, occupy only less than $10\%$.
According to the data above, we created and adapted our DNN model. Fig. \ref{fig: F010} shows our mini experiment results, which are analyzed by Logistic Regression, Decision Tree, Random Forest and SVM algorithms with DNN. We found that accuracy and precision of DNN are more stable. Thus we included more countries in our experiment. According to our experiment, we can perceive except for the arithmetic mean of deep neural network which has reached 85\%. The rests of the Logistic Regression, Decision Tree, Random Forest algorithm only reached between 70\% and 80\%, and SVM only reached 75\% (shown as Fig. \ref{fig: F011}). Moreover, the standard deviation is applied to evaluate the stability of each learning method. Our DNN model has a low standard deviation which indicated that the data points tend to be close to the mean. With the consideration of long term defense and system maintenance of phone scams and the consideration of the detection accuracy of deep neural network, we are pretty sure that the adapted deep learning approach is better than conventional machine learning approaches.

\begin{figure}[hbtp]
	\centering
	\includegraphics[width=3in,height=2.2in]{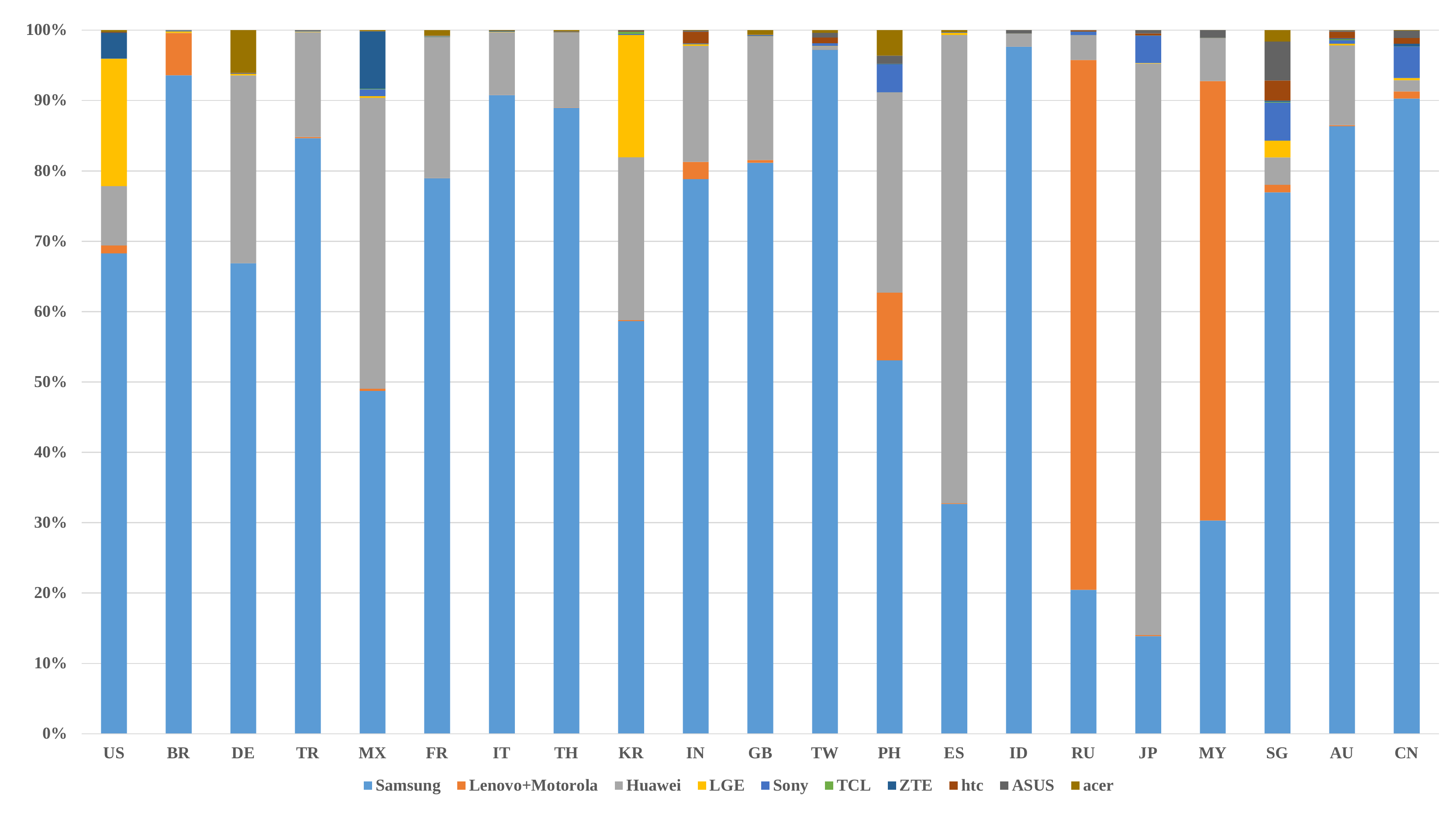}
	\caption{TOP 10 mobile brands' MarketShare.}\label{fig: F013}
\end{figure}

\begin{figure}[hbtp]
	\centering
	\includegraphics[width=3in,height=2.2in]{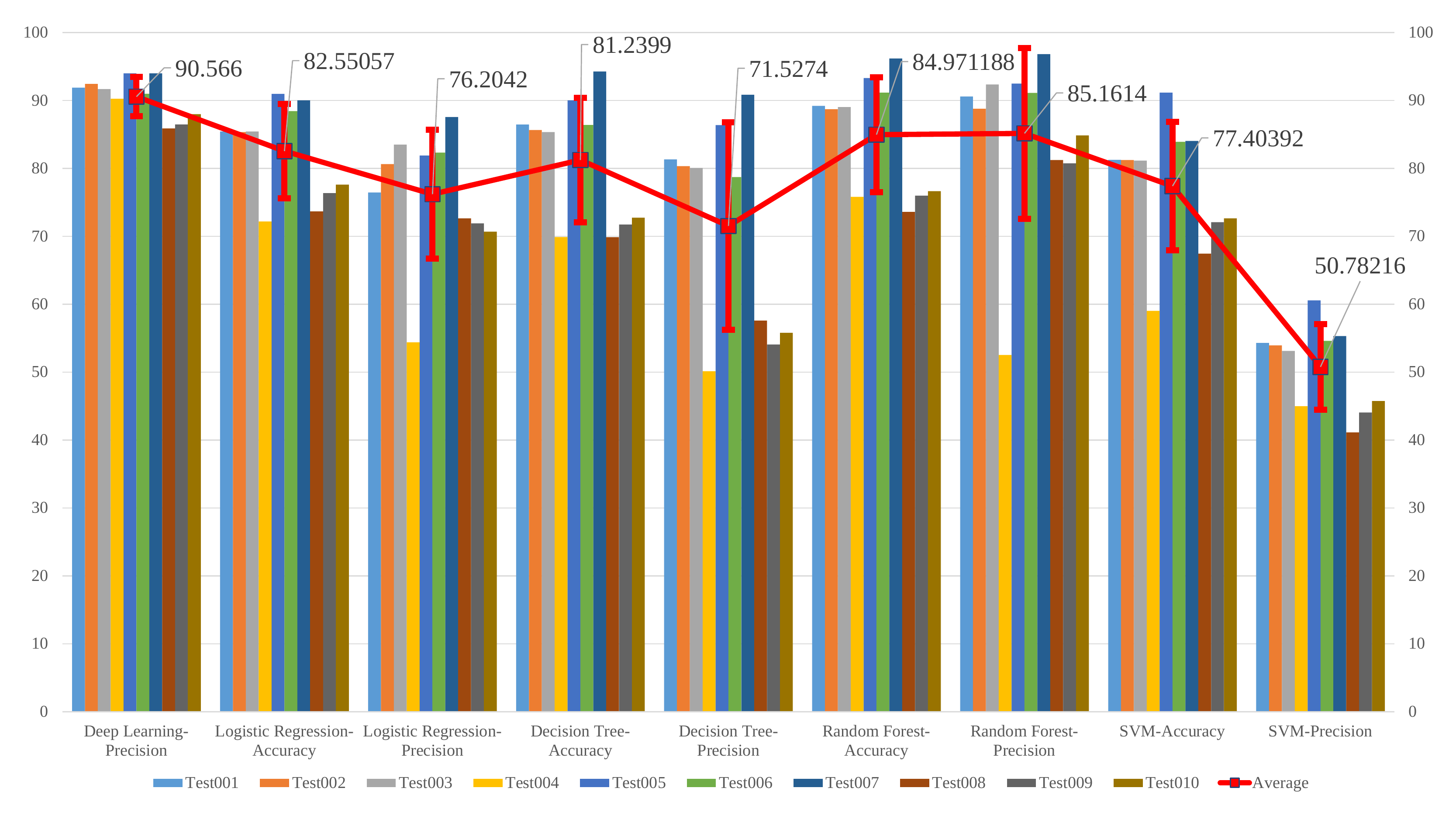}
	\caption{The experiment result.}\label{fig: F014}
\end{figure}

Usually, we can collect 150 thousand URLs including the advertising screenshots and texts per day. According to the data above, we integrated and adapted Google’s deep learning model that has previously trained Inception-v3 for screenshots via TensorFlow. At the same time, due to Android Fragmentation, we found that deceptive ads also has the same Location Based Service (LBS) function of benign advertising. Therefore, we focused on TOP 10 mobile brands’ market share and further discussed the difference between feature extractions, as shown in Fig. \ref{fig: F013}. For instance, Samsung has huge amount of users in the United State, and Huawei also has a lot of users in the Germany and Mexico which cannot be ignored. Fig. \ref{fig: F014} shows our experiment results, which are analyzed by Logistic Regression, Decision Tree, Random Forest and SVM algorithms with Inception-V3. According to our experiment, we can perceive except for the arithmetic mean of Inception-v3 that has reached 90\%. The rests of the Logistic Regression, Decision Tree, Random Forest algorithm only reached between 70\% and 85\%, and SVM only reached 50\%. Moreover, the standard deviation is applied to evaluate the stability of each learning method. The Inception-V3 had a low standard deviation which indicated the data points tend to be close to the mean. With the consideration of long term defense and system maintenance of deceptive ads and the consideration of the detection accuracy of Inception-V3, we are pretty sure that the adapted deep learning approach is better than conventional machine learning approaches. Furthermore, our field test shows that Kaspersky, AVG, Avast, ESET and Chrome failed to detect the deceptive ads.

\section{Conclusion and Future Work}

\begin{figure}[hbtp]
	\centering
	\includegraphics[width=3.3in,height=2.6in]{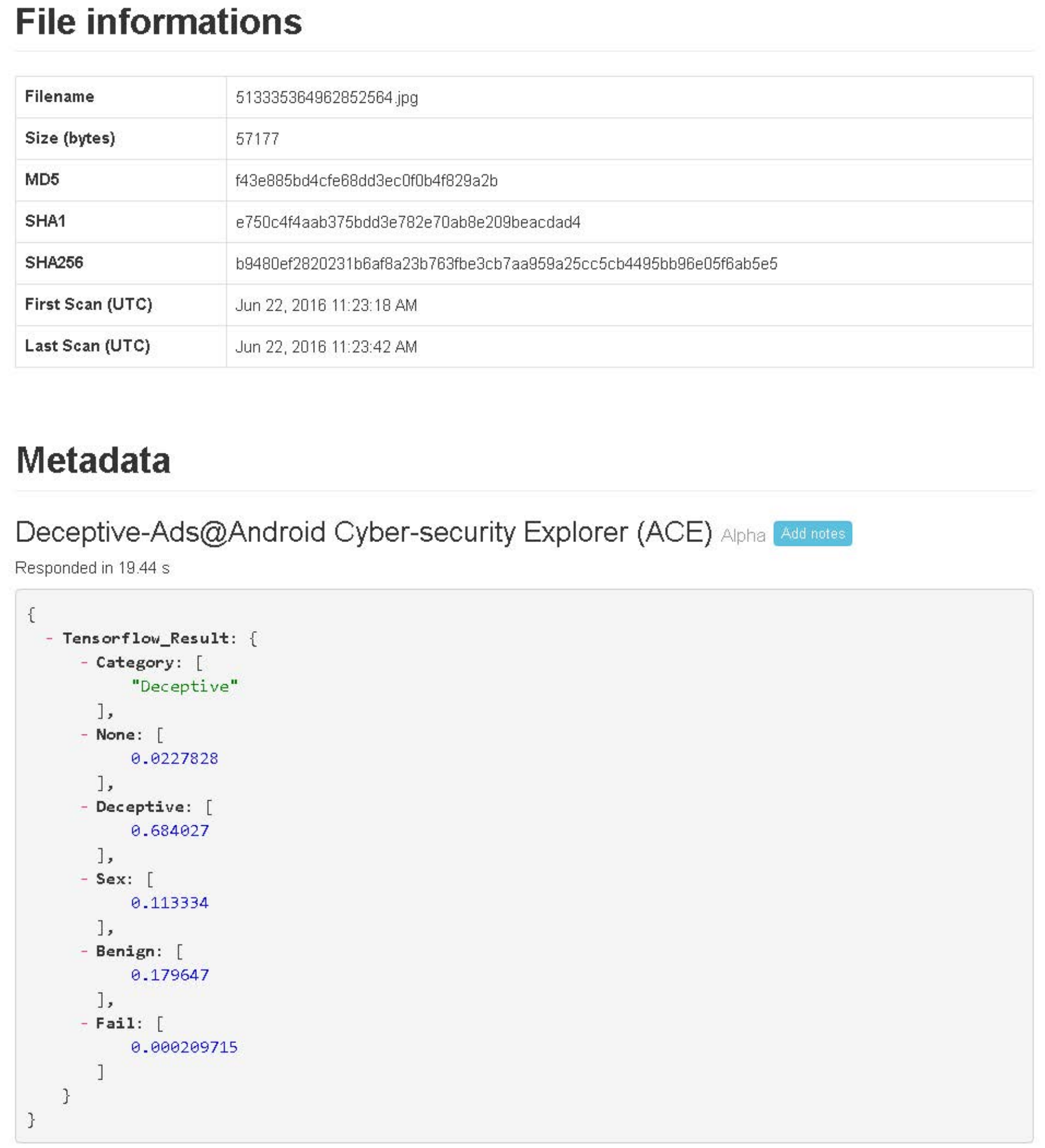}
	\caption{The analyzed results of our proof-of-concept system.}\label{fig: F015}
\end{figure}

The proposed proof-of-concept system was tested in our internal environment. Fig. \ref{fig: F015} shows the analyzed result, respectively. The results show that our detection system works very well to detect phone scams and deceptive ads. Also, we have published the system to our core product to provide convenient usage scenarios for end-users or enterprises. The future work is the improvement of our deep learning model and to reduce the complex task and to train a high performance system for the phone scams and deceptive ads from a huge amount of computation burden.

\section*{Acknowledgment}

Many thanks to the Leopard Mobile, Cheetah Mobile and Security Master (a.k.a CM Security, an Android application) for their valuable dataset.

\end{document}